# Interactions of exotic particles with ordinary matter

Ionel Lazanu [a] and Sorina Lazanu [b1],

[a] *University of Bucharest, Faculty of Physics, P.O. Box MG-11, Bucharest-Măgurele, Romania*
[b] *National Institute of Materials Physics, P.O. Box MG-7, Bucharest-Măgurele, Romania*

**Abstract**

Weakly interacting massive particles (WIMPs) and strangelets are two classes of "exotic" particles not yet discovered, and in agreement with theoretical scenarios most probably produced in different early stages of evolution of the Universe. Some peculiarities of their energy loss in the electronic and nuclear interactions with ordinary matter are investigated. For the direct detection of WIMPs the signals produced by the stopping of recoils in matter are used for their identification. The influence of the orientation of the recoil in respect to crystal axes for crystalline silicon (as material for detectors) is analysed as average quantities: energy loss, and as transient thermal effects. For strangelets, the mechanisms of picking-up neutrons during their penetration into matter and the effects on electronic and nuclear stopping are considered. The clarification of the aspects related to the stopping of these hypothetical particles in matter will permit a better interpretation of some experimental results and could also contribute to the search for new techniques or materials for their detection, if they exist.

## 1. Introduction

New evolutions in astroparticles, cosmology and particle physics open the way to new theoretical predictions about the existence of new states of matter, most of which were produced in the early stages of the evolution of the Universe. If these new constituents of the matter are stable, have a very low probability of interaction and exist in high densities in the Universe, they could have survived until today and could be viable candidates for dark matter (DM).

In the present paper, the peculiarities of the electronic and nuclear energy loss in different targets for distinct types of "exotic" particles: a) WIMPs (Weakly Interacting Massive Particles), and b) strangelets are investigated.

---

[1] Address : 105bis Atomistilor Str, Magurele, 77125 Romania ; e-mail: lazanu@infim.ro; phone: +40-213690170; fax: +40-213690177

In the next section, the characteristics of each class of these "exotic" particles are reviewed, and peculiarities of their interactions with ordinary matter are considered. These exotic particles represent two distinct classes because their dominating interactions are the weak and strong one respectively, and they could be associated with different moments in the thermal history of the early Universe in the standard image of the unification of interactions. If for a long time these particles were investigated separately, recently it was shown that it is possible that WIMPs create strangelets. In agreement with Pérez-García et. al. [1], self-annihilating of weakly interacting massive particles with masses above a few GeV, accreted onto neutron stars may provide a mechanism to seed strangelets.

Section 3 presents and discusses the results obtained in the frame of the models for each type of exotic particle considered in this paper. The clarification of the aspects related to the stopping of these particles in matter will permit a better interpretation of some experimental results and will contribute to the search for new techniques and for new materials for detection.

## 2. Exotic particles: characteristics and peculiarities of their interaction with ordinary matter

The nature and characteristics of DM is a question of central importance in cosmology, astrophysics and astroparticles. Recent experimental results and observations have greatly expanded the list of candidates and the possible signatures of dark matter. A summary of dark matter particle candidates, their properties, and the potential methods for their detection was recently given by Feng [2].

At present, the space missions of AMS-02 [3] and Pamela [4] Collaborations represent a state-of-the-art of the direct investigation of the cosmic radiation, antimatter and some hypothetical exotic components of the Universe outside the atmosphere of the Earth, complementary to the extended underground DM searches (see for example [5, 6].

### 2.1 WIMPs

From all DM candidates, WIMPs are the most studied, are found in many particle physics theories, have naturally the correct relic density, and could be detected in many ways. The candidates for WIMPs have the mass in the range $m_{weak}$ = 10 GeV – 1 TeV, weak interaction is dominant, have tree-level interactions with the W and Z gauge bosons as well as with the gravitational one. No interactions mediated by gluons or photons are permitted.

The peculiarity of the energy loss by a WIMP is the transferred energy to recoil after, most probably, a singular scattering process produced in the detector.

Starting from the general principles of WIMP detection, it must be mentioned that in the literature a classification of the detection methods in direct and indirect exists. In fact, the use of the terminology of direct detection has a specific significance for DM physics, because the very low probability of interaction of WIMPs does not favour them to produce a direct signal in the detector. Hence, in the direct detection technique, the WIMP interacts with a nucleus of the detector and this interaction generates a nuclear recoil whose subsequent interactions are associated with the primary WIMP. Due to the small interaction rate, most probably only a single primary interaction is produced, and for the same reason it is expected the WIMP interaction could be produced anywhere in detector, with a uniform distribution in the depth of the sample. In contrast to this, the interactions induced by "normal" ions with penetration length smaller than the detector size do concentrate near the surface of the detector. The recoil spectrum has an approximately exponential shape. If the spin – (in)dependent interaction is used in the detection, different dependencies of the target mass number must be considered [7]. The indirect detection of WIMPs, which will not be discussed here, consists in observing fluxes of secondary particles created by WIMP annihilation in various astrophysical objects.

Typically, the WIMP velocity relative to the detector is of the order of the galactic rotation velocity, i.e. around 200 km/s, and thus the kinematics is non-relativistic. The main interaction is elastic diffusion of the WIMP on the target nucleus. The recoil energies are in the range 1 up to 100 keV. For WIMP searches, this requires the use of low threshold detectors, which are sensitive to individual energy deposited of this order of magnitude.

In elastic scattering an incident particle, ion or recoil, transfers an amount of its energy to an atom in the target (elastic energy transfer). If the projectile is nonrelativistic as supposed, the transferred energy ($T$) is $0<T\leq\gamma E$, where $\gamma=4m_p m_t/(m_p+m_t)^2$, $m_p$ and $m_t$ being the projectile and target mass respectively, and $E=(1/2)m_p v^2$ is the kinetic energy of the projectile. The transferred energy is determined by the forces between the particles; typically a dependence $(d\sigma/dT)\sim T^{-1}$ is obtained. Due to screening by the electrons, weaker collisions with nuclei are produced and a lower energy is transferred, $(d\sigma/dT)\sim T^{-1+1/n}$, where $n \to \infty$ for strong screening. The screening effect is used to define the cut-off on the energy transfer [8]. Usually the Bohr adiabatic criterion is used. For large $E$ and close collisions with the nucleus, the elastic energy transfer $T$ could be greater than the binding energy of nucleons in the nucleus and thus nuclear reactions could be induced. Momentum transfer collisions produce a cascade of other collisions; all these particles are referred as recoils. In solid materials, defects could be produced if the energy transferred to the struck target atom is greater than the threshold energy for displacement. The electrons in atoms screen the nuclear charges, but also they could be considered as targets and electronic energy loss is produced in competition with nuclear processes. A unitary theory does not exist, but major results could be cited covering different energy ranges [9]. Fast projectiles produce electronic excitations and ionizations if their energy is higher than 1keV/amu. The mean free path is $[n_B(d\sigma/dT)dT]^{-1}$ with $n_B$ the concentration of scattering centres for both classes of interactions.

Different targets used as detectors have crystalline structure and the possible channelling effect was analysed in the literature – see for example Refs. [10 – 13], as a way for the new class of directional studies for WIMPs detection, but the effects of the disorientation of the semiconductor material were not considered.

As specified before, typical information associated with the WIMP interaction is given by the nuclear recoil, the effects depending on mass, electrical charge, kinetic energy and target properties (density, binding energy, scintillating properties, etc.). For detection, signals from ionization, scintillation, heat and calorimetric information could be used directly. An alternate method is to use information in correlation, exploiting different technologies and materials (heat-and-ionization in germanium or silicon crystals at cryogenic temperatures, heat-and-scintillation, ionization and scintillation in noble gases (Ar and Xe) - using or not dual-phase noble gas TPC for example, or superheated droplet technique. Independently on the detection method, the effect of the recoils is disorder production, either transient or permanent. At energies of the recoil high enough, only electronic stopping is important, and with the decrease of the energy, near the end of range of the recoil, nuclear stopping, caused by particle - atomic collision, dominates. In the process of stopping, a local heating effect is produced and the evolution in space and time was previously modelled by the authors [14 – 16]. Phase transitions or local permanent structural modifications of the detector material are possible in particular cases and could be used as a supplementary indirect signal for detection.

## 2.2 Strangelets

The normal nuclear matter is composed by up and down quarks. From theoretical considerations, it was suggested [17] that at densities slightly above nuclear matter density, a strange quark matter composed of up, down, and strange quarks in roughly equal numbers could exist and it could be absolutely stable. Contrary to WIMPs case, for strangelets the dominant interaction is strong. Several models of strange quark matter exist in the literature, see for example [18, 19], suggesting that the strange quark matter could be energetically favourable depending on the values of relevant parameters. Masses per baryon below the mass per nucleon in nuclei (by assumption of stability), and a very low charge-to-mass ratio compared with ordinary nuclei are common to all types of strangelets. In fact it is expected that strangelets have a nucleus structure with a high $A/Z$-ratio compared to nuclei, that they could be neutralized by electrons and form unusual ions or atoms. This is the most important experimental signature for strangelets detection supposing their stability. In agreement with the model for the strangelets with colour-flavour locked quark matter, a $Z = 0.3A^{2/3}$ dependence is predicted and for the strangelets without pairing ("ordinary" strangelets) [20], dependencies as $Z = 0.1A$ for $A \ll 700$ and $Z = 8A^{1/3}$ for $A \gg 700$ are predicted.

The estimation of the minimum value of A for long-lived strangelets is model-dependent, but typical minimum $A$ values are in the range $A_{min} \approx 10–600$ (up to Z≤70, or $A \approx 680$ with $Z \approx 105$ in the other case - colour-flavour locked strangelets) [21]. These predictions are all consistent with the result $Z \approx 0.1 A$ from the literature.

Strangelets with 'magic' numbers of quarks are found for bag constant values $145 \leq B^{1/4} \leq 170$ MeV. If smaller strangelets with mass number smaller than the minimum $A$ are created, they will decay rapidly [22].

Strangelets are expected to possess a small positive electric charge. It is also possible they are neutral, if the ground state composition consists of equal numbers of quarks of the three types of quark flavours, which is the most favourable state. For strangelets a variety of experimental searches have been performed including heavy ions activation [23], emulsion chambers [24], mass spectrometry and accelerator as a mass spectrometer [25] for example as terrestrial experiments, or in space (using balloon in the Earth's atmosphere or terrestrial missions as AMS -01, or placed in the International Space Station: PAMELA and AMS). Some overviews of the past searches and a more detailed description of recent ones were given in Refs. [26 – 29].

Some singular events observed in different experiments are considered as candidates for strangelets. Centauro events were observed very deep in the atmosphere (at about 500 g/cm$^2$) and contain probably about 200 baryons [30]. The so called "Saito events" have $Z \sim 14$ and $A \sim 350$ and $A \sim 450$ respectively [31]. Price [32] reported an event with $Z = 46$ and $A > 1000$. The Exotic Track event [33] has been produced after the respective projectile has traversed ~ 200 g/cm$^2$ of atmosphere. The most interesting candidate has $Z/A = 0.114 \pm 0.01$ and a kinetic energy of 2.1 GeV, assigned to $^{16}$He [34]. The background from ordinary nuclei was estimated to be less than $10^{-3}$ events. This event was identified by the Alpha Magnetic Spectrometer (AMS-01) experiment, when the detector was flown on the space shuttle Discovery during flight STS–91 (1998) in a 51.7 degree orbit at altitudes between 320 and 390 km [35]. A second candidate of AMS -01 Collaboration was reconstructed as having a nuclear charge of Z =+8 and a mass $A = 54^{+8}_{-6}$, which we will denote as $^{54}$O. This result is cited from Ref. [36].

Some authors assume that the strangelets which come to the upper layer of the atmosphere have baryon number *A* of the order 1000 or more and these candidates move in large atmospheric depths requiring unusual penetrability which means that their cross sections should be very small and hence a geometric size much smaller than typical nuclear size. In this model, it is supposed that for the events measured the initial masses decrease rapidly due to their collisions with air molecules. In the work of Paulucci *et al.*, [37] different mechanisms of interactions of strangelets with ordinary matter are discussed.

Alternatively, there exists a model based on the following hypothesis: in a typical interaction between a strangelet and the nucleus of an atmospheric atom, it is more probable for the strangelet to absorb neutrons from the colliding nucleus, and as a result the mass of the strangelet increases in every collision and it becomes more tightly bound [38, 39]. This scenario is in accordance with the events observed by AMS. If this hypothesis is accepted, neutron capture is produced in different traversing media (atmosphere, material of the detector system), and is more significant in materials with higher density. These aspects are discussed in this paper.

Following the scenario proposed by Banerjee *et al.* in Refs. [38] and [39] we consider that the strangelet, during its penetration through the detector material picks up mass (neutrons) from target atoms. For the concrete calculations we suppose that silicon is the detection material. The rate of change of mass of the strangelet, $m_{str}$, with respect to the penetration depth *x* is given by the equation:

$$\frac{dm_{str}}{dx} = \frac{fn_n}{\lambda} = fn_n n_t If \quad (1)$$

where $\lambda$ is the mean free path of the strangelet into the material, between two successive interactions, $m_n$ is the mass of the neutrons in the target nucleus, *f* a geometric factor which describes the fraction of neutrons absorbed from the target nucleus in the interaction, $n_t$ is the concentration of atoms in the target, and $\sigma$ is the interaction cross section.

The factor *f* is the ratio between the neutrons participating in the interaction with the strangelet and the total number of nucleons in the target nucleus. An analytical approach for the calculation of this factor is given in Ref. [40], and the formulae from this reference were used here. The scattering cross section for neutron capture is given by:

$$\sigma = \pi(R_{str} + r_t)^2 \quad (2)$$

where $r_t$ is the radius of the target nucleus, and $R_{str}$ is the strangelet radius, having an approximate dependence of its mass number $A_{str}$ [41] as:

$$R_{str} = 0.9973 \times A_{str}^{1/3}$$

The strangelet is stopped due to its mass increase, to its interaction with the electrons and nuclei of the target:

$$m_{str}\frac{dv}{dt} = -v\frac{dm_{str}}{dt} - S_{el} - S_{nucl} \quad (3)$$

where $v$ is the mass of the strangelet, $S_{el}$ and $S_{nucl}$ its electronic and nuclear stopping powers in the material of the target respectively.

For high energies of the strangelet ($v>2v_{Bohr}$), the relativistic Bethe formula for the energy loss per distance travelled is used:

$$S_{el} = \frac{4\pi}{m_e c^2} \cdot \frac{n_{Si} Z_{Si} Z_{str}}{\beta^2} \cdot \left(\frac{e^2}{4\pi\varepsilon_0}\right)^2 \cdot \left[\log\left(\gamma^2 \frac{2m_e c^2 \beta^2}{I}\right) - \beta^2\right] \qquad (4)$$

where $\beta = v/c$, $E$ is the energy of the strangelet, $Z_{str}$ and $Z_t$ are the charge numbers of the strangelet and target respectively, $m_e$ is electron's mass and $I$ is the mean excitation potential of the target. In the following calculations the target has explicitly/ considered to be silicon.

At low energies of the strangelet, Lindhard's formula [42] for the electronic stopping power proportional to the velocity is used:

$$S_{el} = k \frac{Z_{str}^{7/6} Z_{Si}}{\left(Z_{str}^{2/3} + Z_{Si}^{2/3}\right)^{3/2}} \cdot \left(\frac{E}{m_{str}}\right)^{1/2} \qquad (5)$$

($k=3.83\ 10^{-15}$, $S_{el}(E)$ expressed in eVcm²/atom). As a first approximation, only these two velocity (energy) regions are considered, disregarding the transition intermediate energy region, whose analytical approach is more difficult.

For the nuclear stopping power, we used the Ziegler-Biesack-Litmark formula [43]:

$$S_{nucl}(E) = g \frac{Z_{str} Z_{Si} A_{str}}{(A_{str} + A_{Si}) \cdot (Z_{str}^{0.23} + Z_{Si}^{0.23})} S_n(E) \qquad (6)$$

($g=8.462\ 10^{-15}$ when $S_{nucl}(E)$ is expressed in eVcm²/atom).

The nuclear stopping power in formula (6) is defined in respect to a reduced nuclear stopping cross section $S_n(\varepsilon)$:

$$S_n(\varepsilon) = \frac{0.5 \log(1+1.1383\varepsilon)}{(\varepsilon + 0.01321 \cdot \varepsilon^{0.21226} + 0.19593 \cdot \varepsilon^{0.5})} \qquad (7)$$

which in its turn depends on the reduced energy ε:

$$\varepsilon = \frac{32.53 \cdot A_{Si}}{Z_{str} Z_{Si} (A_{str} + A_{Si}) \cdot (Z_{str}^{0.23} + Z_{Si}^{0.23})} E \qquad (8)$$

Equations (1), (3), together with the definition of the velocity:

$$v = \frac{dx}{dt} \qquad (9)$$

represent a system of ODEs for the unknowns $A_{str}$, $v$ and $x$, with the initial conditions: $A_{str}(0) = A_0$, $v(0) = v_0$, and $x(0) = 0$.

The system has been solved numerically for $A_{str}(t)$, $v(t)$ and $x(t)$, from $t = 0$ up to the stopping time $t_{st}$, which is defined from the condition $v(t_{st}) = 0$.

Only neutron capture is considered here and the possibility of mass increasing due to proton capture is disregarded.

## 3. Results and discussions
### 3.1 WIMPs

As a result of the elastic interaction of a WIMP from the galactic halo with the target nuclei of the detector, a small amount of energy, less than 100 keV, is transferred to the recoil nucleus. In Figure 1, the recoil energy dependence on the WIMP mass for different crystalline targets, semiconductors and solid noble gases – new promising materials, is presented. For all the curves, the velocity of the WIMP is taken 230 km/s, and an average of recoil energies over scattering angles is considered. It could be observed that in all targets, the range of recoil energies is very narrow and closes in a common range for WIMP masses up to 70 GeV/$c^2$, and in all cases is less than 40 keV.

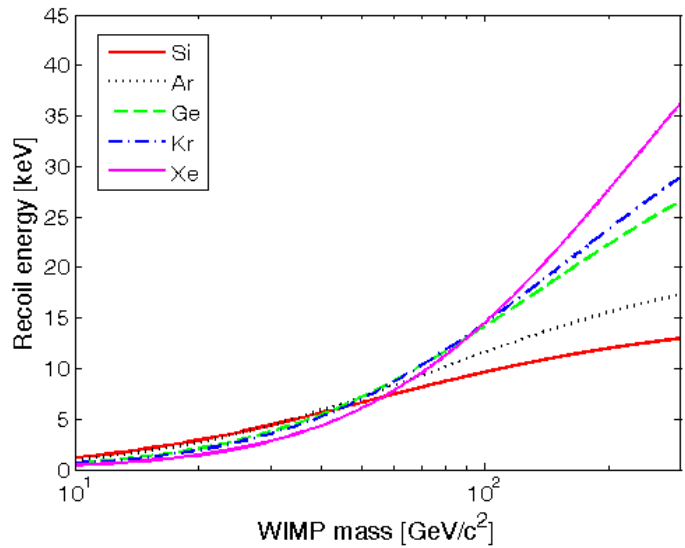

Fig 1: Recoil energy as a function of the WIMP mass for elastic scattering averaged over angles, for the following target materials (from bottom to top at the right extremity): Si, Ar, Ge, Kr and Xe.

The recoil produced in the interaction with the WIMP is stopped in the target due to its interactions with the electrons and nuclei. The dependence of the electronic and nuclear energy loss on the WIMP mass are presented in Figure 2, considering as targets Si, Ge, and solid noble gases Ar, Kr and Xe, which are promising detector materials. The energy losses are from SRIM [44]. For all materials, the nuclear energy loss is higher.

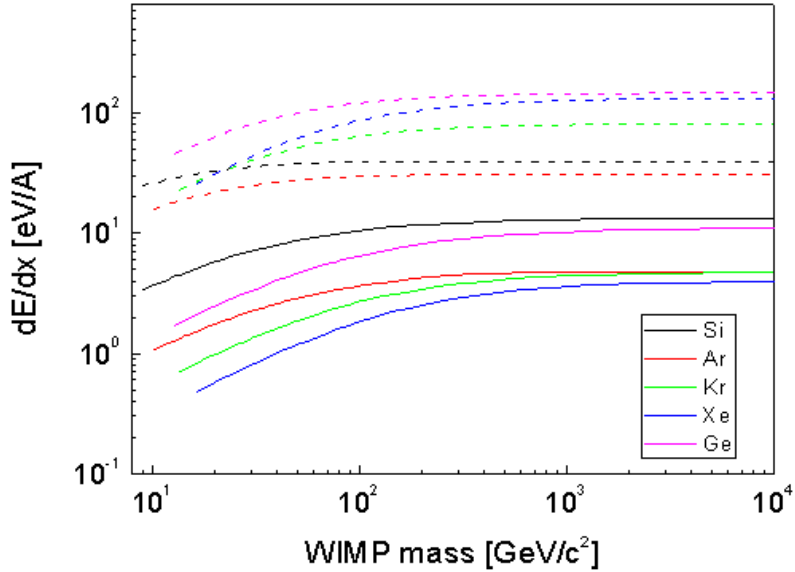

Figure 2: Electronic (continuous line) and nuclear (dashed) energy loss of a selfrecoil in Si, Ge, solid Ar, Kr and Xe respectively, produced by a WIMP of 230 km/s, as a function of the mass of the WIMP.

To illustrate the influence of space arrangement of the atoms in the crystal on the process of penetration of the self-recoil in the target, as well as the influence of the relative orientation of the recoil in respect to the crystalline planes, we used the binary collision code *Crystal-transport and range of ions in matter* (Crystal-TRIM) [45]. The penetration of 500 Si ions of 15 keV kinetic energy, incident along the <100> and <111> channelling directions, and at different disorientations in respect to them, defined by the angles $\theta$ and $\varphi$ was simulated (see Ref. [46] for the definition of the angles).

The dependence of the electronic and nuclear energy loss on the penetration depth obtained with Crystal TRIM was compared with the corresponding one obtained with SRIM, which disregards the crystalline structure of the target. The results are displayed in Figures 3 a and b. As expected, the penetration depth is significantly higher if the recoil follows the directions of crystalline axes, and as a limiting case, channelling is produced.

During slowing-down, the energy of the primary recoil created by the WIMP interaction is imparted to both the electronic and lattice (nuclear) subsystems of the target. A local, transient, heating effect is produced, and it is treated here in the frame of a spike model with two sources, corresponding to the electronic and nuclear energy losses respectively [15]. After the processes by which the recoil loses its energy in the medium, the two subsystems have different temperatures and are coupled through a term that is a measure of the energy exchange, the electron-phonon coupling. The processes are studied in a thin layer, perpendicular to the track, and a cylindrical symmetry is considered — see in ref. [16] a discussion on the applicability of the cylindrical and spherical spike.

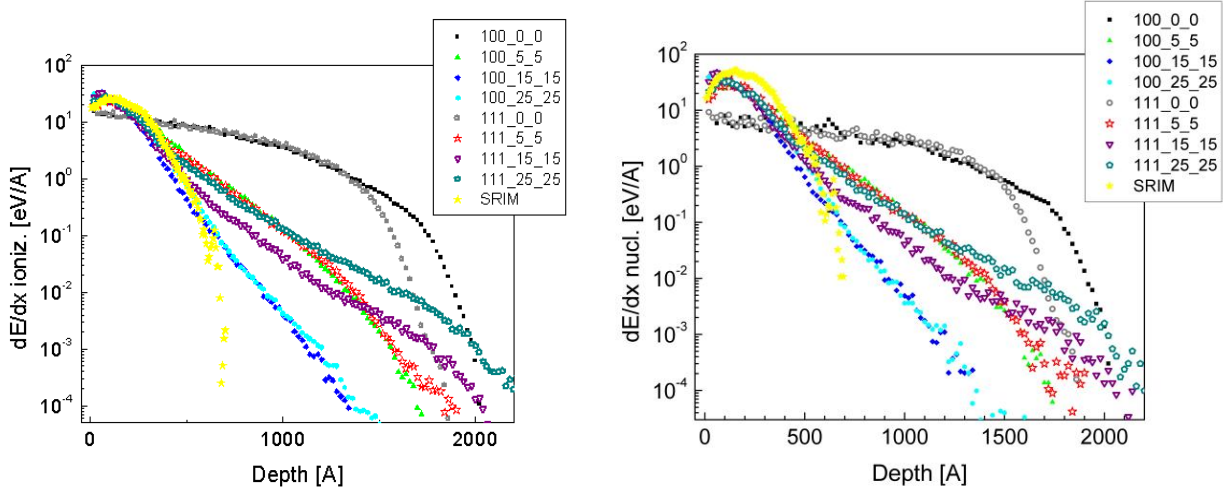

Figure 3: Depth dependence of the electronic (left) and nuclear (right) energy loss of a 15 keV selfrecoil in silicon, with the indication of the tilt and rotation angles of the direction of the particle for crystalline targets.

The localized regions of the medium characterized by departure from equilibrium due to the energy transfer from the projectile toward electrons and nuclei respectively are generally different because the mechanisms of interaction and the kinematics are distinct. The dependence of the electronic and atomic temperatures on distance to the recoil trajectory and time elapsed from its passage, obtained as solution of the system of coupled differential equations from Ref. [14] are represented for the recoil which follows the 100 direction, and for a direction which differs from it by 15 degrees in the angles $\theta$ and $\varphi$, in Figures 4 and 5 respectively. The comparison of these figures gives an idea on the effect of the anisotropy of the crystal, hence of the dependence of the heating effect produced by a 15 keV Si selfrecoil on its direction of motion in the crystal.

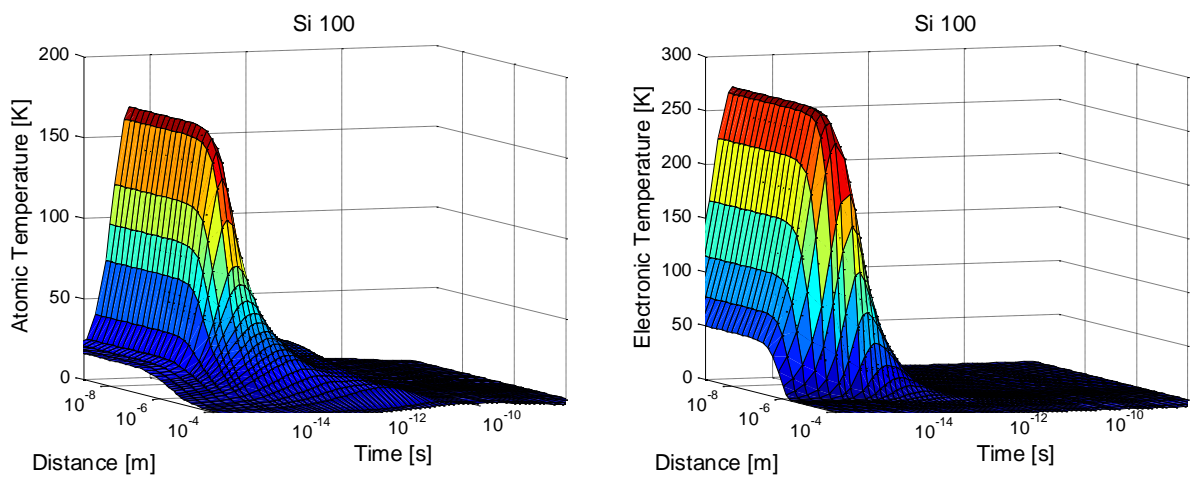

Figure 4: Lattice (left) and electronic temperature (right) dependence on distance and time for a Si selfrecoil of 15 keV oriented along the 100 direction.

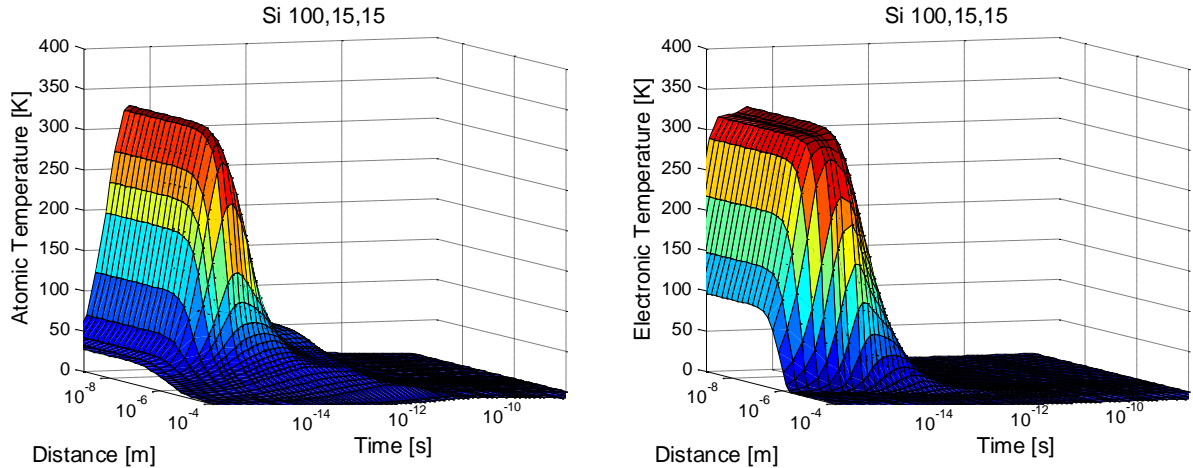

Figure 5: Lattice (left) and electronic temperature (right) dependence on distance and time for a Si selfrecoil of 15 keV oriented along a direction desoriented in respect to 100 by 15 both on θ and φ.

Both the increase in the atomic and electronic temperature produced by the channelled selfrecoil are lower in respect to that produced by the disoriented one, due to the fact that the ion going straight on the <100> direction loses energy (both to the electronic and atomic subsystems) at a considerably lower rate. The time dependence of both the electronic and atomic temperature in the case of the channelled ion is much sharper. The increase of the temperature in the electronic system is faster for the disoriented recoil, but the decrease is slower. The coupling between the two subsystems is visible especially for the disoriented recoil: in the atomic temperature space –time evolution it appears as a broad shoulder in the distribution.

These results put clearly in evidence the importance of the correct knowledge of relative orientation recoil – crystalline material in the detection process and could be a useful guide for experiments.

### 3.2 Strangelets

The evolution of strangelets during their penetration through atmosphere was modelled, and the effect of picking-up neutrons on their mass was analysed for a large range of initial mass numbers, from 20 up to 500, and for different initial velocities. For the density of Earth's atmosphere, the information from [47] was used, obtained in the empirical and global model NRLMSISE-00, with input parameters considered in respective figure. The nitrogen and oxygen have been considered as the only components of the atmosphere composition. The increase of the mass number for strangelets penetrating terrestrial atmosphere was obtained as numerical solution of the system of coupled ODEs eqs (1), (3), and (9).. The results are presented in Figure 6. In all cases a *Z=0.1×A* dependence between mass and charge numbers was supposed.

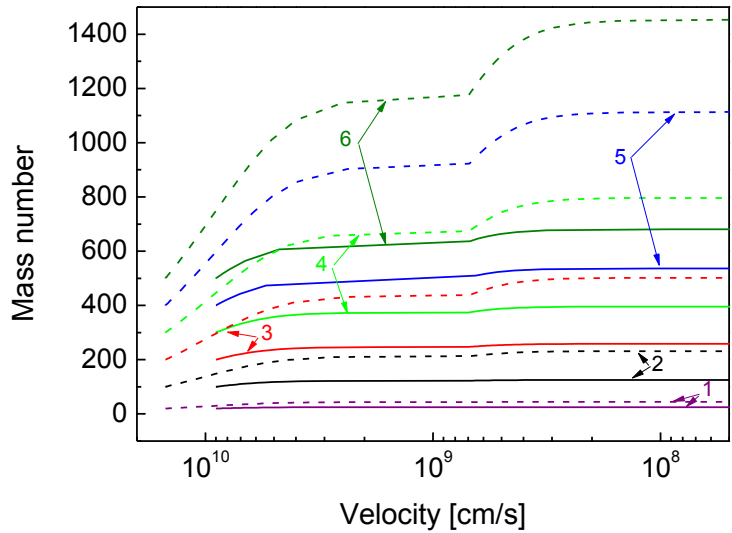

Figure 6: Mass number dependence on the velocity for strangelets penetrating terrestrial atmosphere and having initial velocities: 0.3×c (continuous line) and 0.5×c (dashed line) respectively, for the following initial mass numbers: 20; 100; 200; 300; 400; 500 (curves 1 up to 6)

The evolution of the particles associated with the events reported in the literature and mentioned in the previous section was modelled in different targets, of interest for detection, also by solving the system of ODEs. In figure 7, the mass of the strangelet as a function of the penetration depth in different targets is presented. Both the strangelet (2,16) and (8,54) are taken as examples, both with initial velocities 0.45 c.

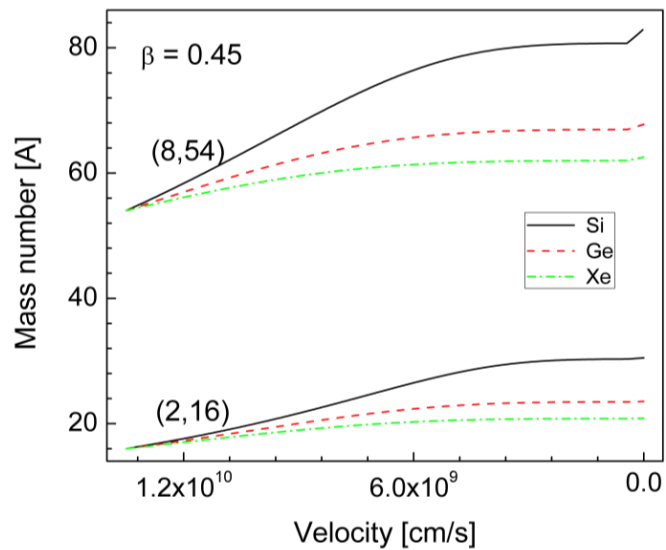

Figure 7: Velocity dependence on the mass number of the strangelets (2,16) and (8,54) in Si, Ge and Xe respectively.

In Figures 8 a) and b) the decrease of the velocity, due to the electronic and nuclear stopping, and to the increase of the mass is presented in different targets as a function of the penetration depth, for two initial velocities.

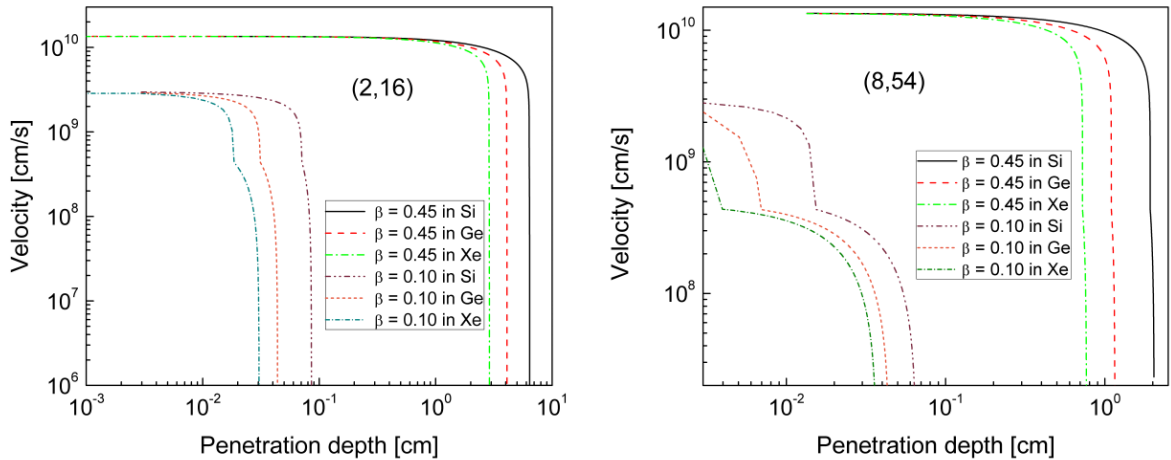

Figure 8: Velocity dependence on the penetration depth for strangelets (a) (2,16) and (b) (8,54) in Si, Ge, and Xe for two initial velocities of the particles, corresponding to β = 0.1 and 0.45 respectively.

The increase of the strangelet's mass during its penetration in the target in represented in Figures 9 a) and b). In the first one, the increase of the mass of both (2, 16) and (8, 54) strangelets in three targets: Si, Ge and Xe is represented, while in the second the same dependence is presented for the strangelet (2, 16) with different initial betas. It appears that there exists a "universal" curve for each strangelet's dependence on the penetration depth, the same for different target materials and initial velocities. Depending on the target material, for the same initial velocity of the strangelet, the penetration depth and correspondingly the increase in mass are different.

For different velocities and the same target, dependent on the fastness with which the projectile loses energy, it will stop at different depths. We would like to mention that there is no mass increase neither for the strangelet (2, 16), nor for the one (8, 54) with β = 0.1 in Si because the strangelet cannot capture any neutron. In order to obtain an idea of the shrinkage of the range due to the increase of the mass, the penetration depths of the strangelets (2, 16) and (8, 54) in Si, without neutron capture were calculated: the range for the strangelet (2, 16) varies from 0.82 to 20.29 cm, when β varies between 0.2 and 0.45, and for the strangelet (8, 54), for the same interval of variation of β, the range is in the interval 0.2 – 4.32 cm.

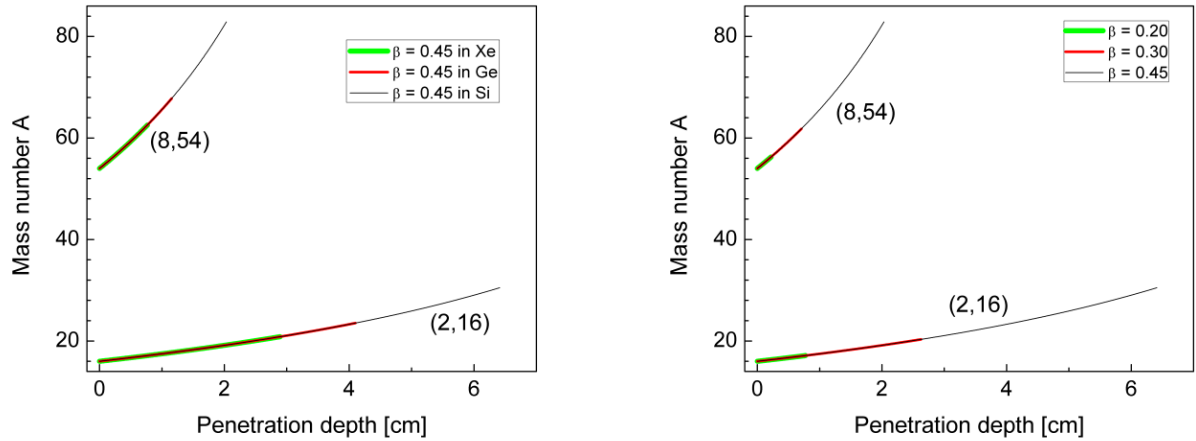

Figure 9: Mass dependence on the penetration depth a) in Si, Ge and Xe for strangelets (2,16) and (8,54) with β=0.45, and b) in Si for strangelets with β=0.45, β=0.30 and β=0.20 respectively

In the identification of the particle, the possibility that the strangelet captures neutrons in the processes of slowing down must be considered. As a consequence, it is possible that the same strangelet to be identified as different particles if it is measured: in space or in higher atmosphere of the Earth; near the surface or underground; or after penetrating massive and great dimension detectors.

## 4. Summary and conclusions

The penetration through (crystalline) targets, potential candidates for detectors, of two classes of exotic particles, relics of the thermal history of the early Universe, and characterised by the weak and strong dominating interactions, was analysed.

WIMPs interaction with matter results in the production of at most a selfrecoil of the detection medium, which in turn loses energy by interaction with electrons and atoms. The analysis of these interactions is the basis of WIMP detection in direct detection methods. The dependence of the electronic and nuclear energy losses of the selfrecoil produced in this way as a function on the WIMP mass, for a given velocity of the particle, was calculated, and the important effect of the recoil orientation in respect to the crystal axes on the stopping was evidenced using Crystal TRIM simulations in silicon. The influence of the recoil orientation on the development of transient effects, treated in the frame of a thermal spike model with nuclear and ionization sources, were also evidenced.

For strangelets, the hypothesis that the projectile picks up neutrons during its penetration in the target was used. The results for the mass and velocity dependence on the penetration depth of some potential strangelet events, of different initial velocities, was analysed as numerical solution of a system of ODEs. The nuclear stopping is very important near the end of range of strangelets, and is taken into account in the present calculations. An ambiguity in the identification of the particle results in the frame of the model, due to the fact that in the detection process the intrinsic characteristics of the strangelet (mass and charge to mass ratio) are modified due to possible capture processes. All these observations are of interest in the search for new techniques and for new materials for detection.

## Acknowledgements

S.L. would like to thank UEFISCDI for support, under Project PNII - IDEI, 901/2008.